\documentclass[journal=apchd5,communication=article]{achemso}

\usepackage[version=3]{mhchem} 
\usepackage[T1]{fontenc}       
\usepackage{caption}
\usepackage{float}
\usepackage{geometry}
\usepackage{natbib}
\usepackage{setspace}
\usepackage{xkeyval}

\author{Ali Eshaghian}
\affiliation{School of Electrical and Computer Engineering, Georgia Institute of Technology, Atlanta, USA}
\author{Sajjad AbdollahRamezani}
\affiliation{Department of Electrical Engineering, Sharif University of Technology, Tehran, Iran}
\author{Ata Chizari}
\affiliation{Department of Electrical Engineering, Sharif University of Technology, Tehran, Iran}
\author{\\Amin Khavasi}
\affiliation{Department of Electrical Engineering, Sharif University of Technology, Tehran, Iran}
\email{khavasi@sharif.edu}

\title 
{Broadband, Polarization-insensitive and Wide-angle Optical Absorber based on Fractal Plasmonics}

\begin{document}


\begin{abstract}
In this paper, a plasmonic absorber consisting of a metal-dielectric-metal stack with a top layer of Sierpinski nanocarpet is theoretically investigated. Such compact absorber depicts broadband angle-independent behavior over a wide optical wavelength range ($400-700$ nm) and a broad range of angles of light incidence ($0-80^{\circ}$). Including several feature sizes, such fractal-like structure shows widely strong extinction ($85-99$\%) response for either transverse electric or magnetic polarization states under normal incidence. Underlying mechanisms of absorbance due to excited surface plasmon modes as well as electric/magnetic dipole resonances are well revealed by investigating electric field, magnetic field and current distributions. The proposed absorber opens a path to realize high-performance ultrathin light trapping devices.
\end{abstract}

\section{INTRODUCTION}

Plasmonic nanostructures and optical metamaterials have emerged as promising candidates for controlling light-matter interactions at subwavelength from terahertz to near infrared. Owing to their interesting physics and exotic properties, engineering spectral properties of light such as absorption, transmission, reflection and emission has been widely realized by manipulating amplitude, phase and polarization of independent electric and magnetic waves responses \cite{atwater2010plasmonics,schuller2010plasmonics,abdollahramezani2015analog,abdollahramezani2015beam,youssefi2016brewster}. Such structures provide improved performance in established technologies such as  photodectors \cite{lopez2013ultrasensitive}, thermophotovoltaic solar energy conversion \cite{avitzour2009wide}, thermal emitters \cite{greffet2002coherent}, and biochemical sensing \cite{liu2010infrared}. 

In all aforementioned modern technologies, perfect black body as the main part which soaks up electromagnetic waves with arbitrary wavelength, polarization and incident angle is much more desirable \cite{kravets2008plasmonic,narimanov2009optical}. As a result, investigating novel approaches to significantly boost the absorption efficiency of black bodies trapping wideband, angle independent and polarization insensitive light has been a major concern for many researchers \cite{aydin2011broadband} . 

Amongst advanced light trapping methods, bodies based on resonant cavities have demonstrated dramatic promise, including Fabry-Perot cavities which can be designed by multilayer thin-film coatings to yield resonant enhancement \cite{watts2012metamaterial}. However, there is a trade-off between their operational bandwidth and absorption efficiency. In addition, their angle dependent optical response necessitate tracking systems in specific applications such as solar cells to maximize achieving energy \cite{cao2010semiconductor}. Using metallic subwavelength structures supporting surface plasmons (SPs), i.e., light induced collective surface electron density fluctuations at the interface between metal and dielectric, have been approved as a promising method for high efficiency energy harvesting systems \cite{watts2012metamaterial,landy2008perfect}. By proper structuring metalic nanoparticles \cite{teperik2008omnidirectional}, grating arrays \cite{bouchon2012wideband} and subwavelength slits \cite{white2009extraordinary}, light can  be concentrated into a thin semiconductor in the form of both localized and propagative surface plasmon polaritons. Although progress in fabrication technologies at nanoscale in the past few years has facilitated the development of optical nanoantennas with various characterization such as size, shape and thickness, sharp resonance response with high quality factor nature of SPs impose narrow operating bandwidths which strongly depends on the incident polarization \cite{cui2014plasmonic}. Accordingly, to achieve a broadband optical response, several pseudo-random, periodic, and aperiodic plasmonic nanostructures have recently been proposed \cite{yan2014lithography,pala2009design,ding2012ultra,zhang2011polarization}.

In this paper, in order to achieve wideband, polarization insensitive, and angle-independent absorber over solar spectrum, we investigate absorption mechanism of an inspired fractal pattern, namely Sierpinski carpet. In scattering theory, reflected, refracted or diffracted phenomena is strongly dependent on the size of scaterrer relative to the wavelength of spectrum. Due to the self-similarity of fractals, different scales within the same structure are replicated resulting in multiple-band spectral response which is useful in broadband applications such as multiband antennas \cite{werner2003overview}. Moreover, fractal structures might have a larger perimeter-to-footprint ratio than its basis shape leading to reduced spaced resonant elements and compact devices \cite{sederberg2011sierpinski}. Our proposed configuration consists of an array of silver nanobricks on top of a thin silicon dioxide platform. The proposed experimentally feasible metallo-dielectric super absorber is capable of absorbing light in the visible spectrum ($400$-$700$ nm) with a simulated average absorption more than $90$\%. Performing three-dimensional (3D) simulations based on finite integration technique (FIT), we theoretically investigate the effect of propagative and localized plasmon modes and localized magnetic and electric dipole resonances on enhancing the efficiency of proposed light trapping structure. These results suggest alternative schemes for design and improvement of subwavelength high performance optical absorbers, sensors and thermal emitters.

\section{STRUCTURE DESIGN}

A schematic image of our proposed broadband absorber is shown in Fig.~\ref{fig1}. As it is observed, each unit cell consists of a three-layer metal-dielectric-metal (MDM) thin-film where a silicon dioxide substrate with thickness $h_{SiO_{2}}$ is settled on a thick silver backreflector with thickness $h_{Ag}$ needed to reflect the incident light back and extends the absorption thickness. The top plasmonic layer is formed in a way that a fractal-like pattern of silver nanobricks is appeared. This pseudo-similar geometrical shape, namely Sierpinski carpet, is constructed with a square of side width $d$ and height $h$ (Fig.~\ref{fig2}(a)); then divided into a grid of $3\times3$ smaller cuboids with width of ${d/{3}}$ and height of ${h/{3}}$, and only the central sub-square is remained(Fig.~\ref{fig2}(b)). The same procedure is then recursively applied to the eight side squares. We continue this process up to the second iteration as shown in Fig.~\ref{fig2}(c) in which all dimensions are compatible with the current nanotechnology fabrication limitations. The dielectric function of silver and refractive index of silicon dioxide are taken from experimental results \cite{johnson1972optical,palik1998handbook}, respectively, and inserted into the CST Microwave Studio commercial software package. Periodic boundary conditions are applied in $x$ and $y$ directions, and the ports at the front and back surfaces of the unit cell are set for the optical ray incidence, reflection and transmission. Furthermore, to consider strong interactions between metal and silicon dioxide, an extremely refined mesh is constructed for the whole structure. 

\section{RESULTS AND DISCUSSION}
 
We first perform numerical computations to investigate the relationship between the absorption spectrum, lattice constant and dielectric spacer thickness. Due to the nature of fractal-like structures, the optimal width of all metal nanocuboids will be achieved automatically by obtaining the optimum lattice constant. Other design parameters are chosen in a manner to be subwavelength and practicable. It is notable that sensitivity of the absorption response to these parameters can also be readily explored.

\begin{figure}
	\centering
	\includegraphics[trim=9.8cm 1.8cm 0cm 0cm,width=16.6cm,clip]{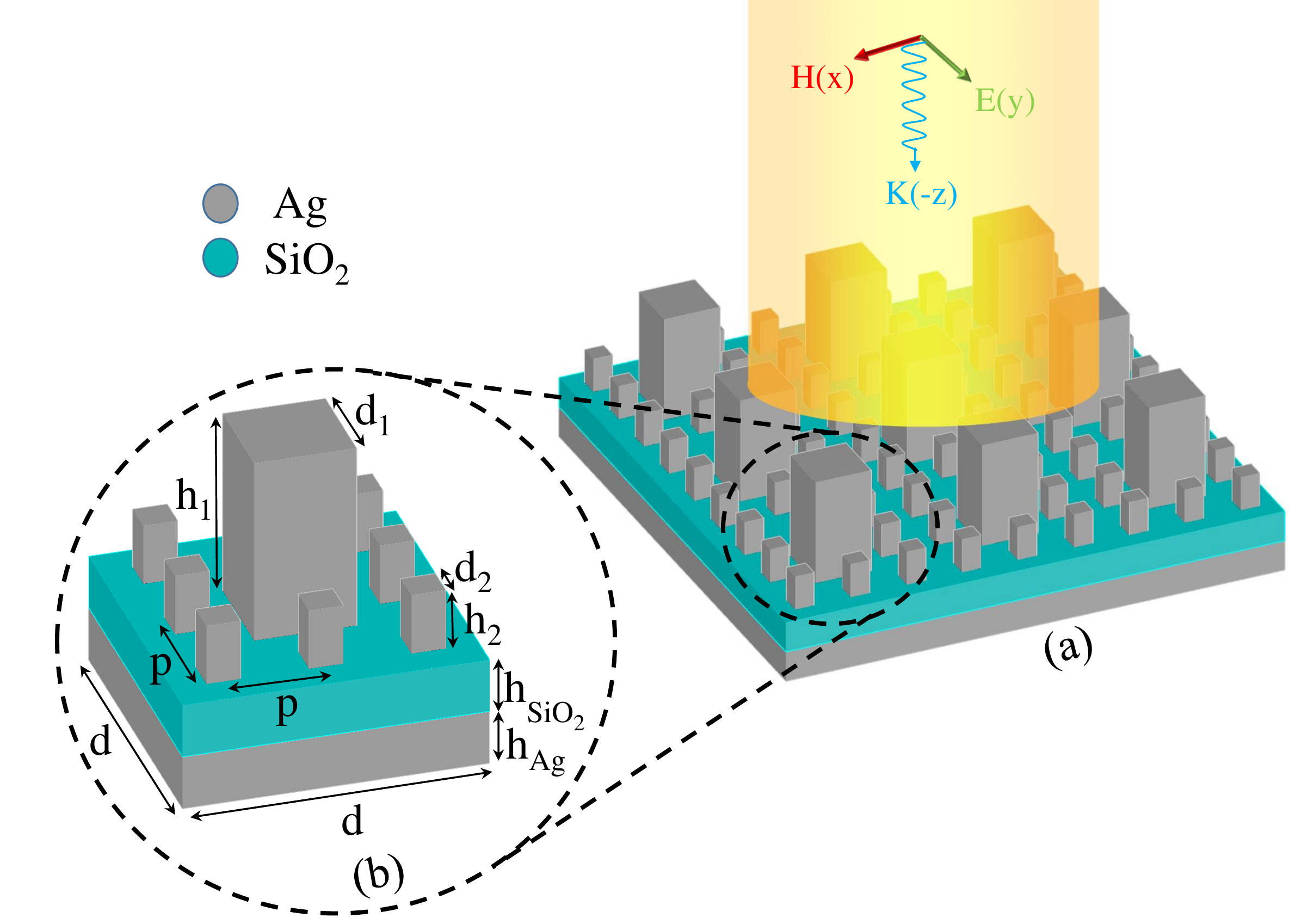}\\
	\captionsetup{justification=justified}
	\caption{(a) Schematic representation of a the proposed MDM absorber with the top layer patterned as metallic Sierpinski carpet. (b) Sketch of a unit cell as a constitutive element.}\label{fig1}
\end{figure}

\begin{figure}
	\centering
	\includegraphics[trim=7.5cm 1cm 0cm 0cm,width=16.6cm,clip]{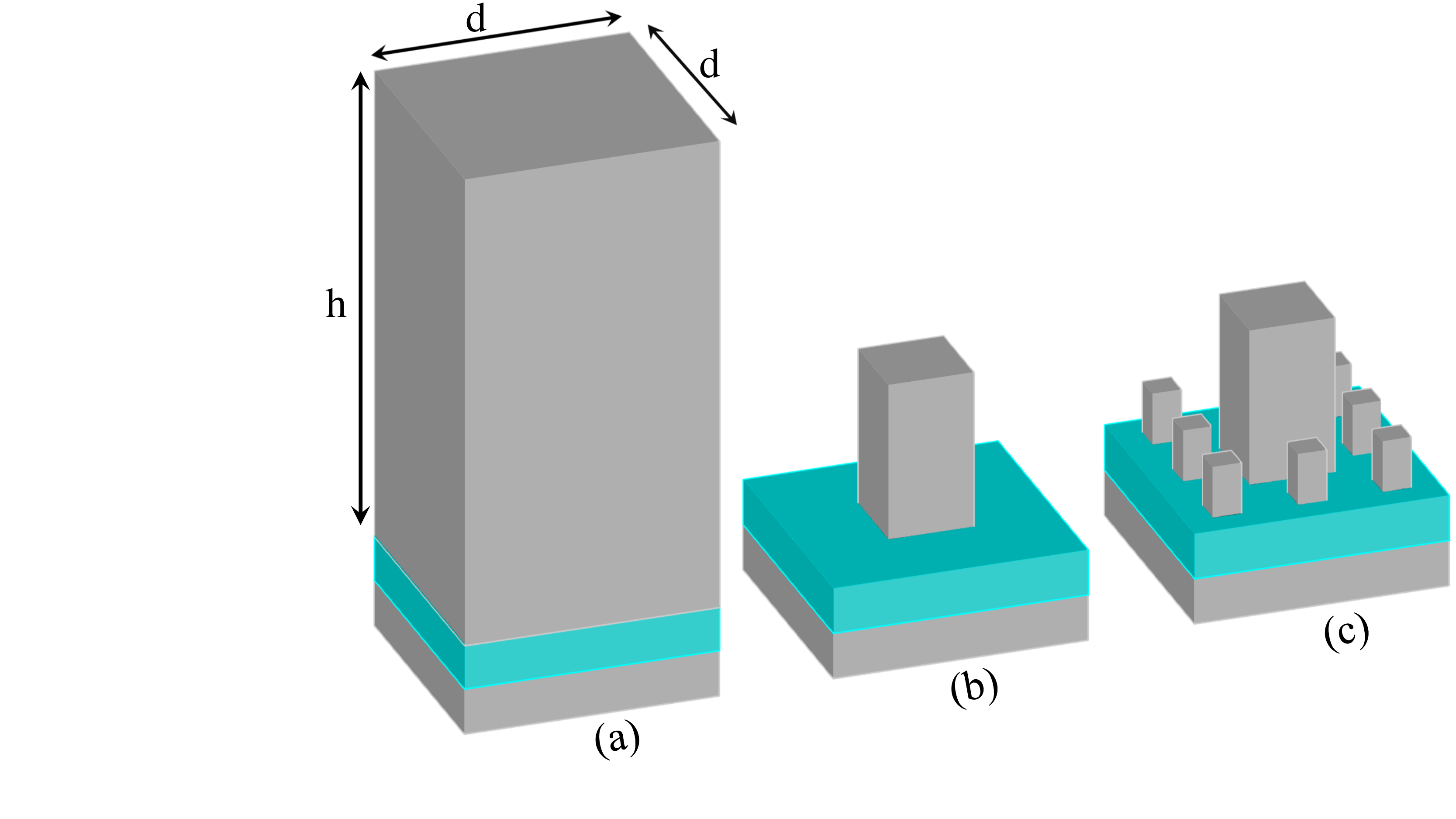}\\
	\captionsetup{justification=justified}
	\caption{Three-dimensional demonstration of (a) zeroth-order, (b) first-order, and (c) second-order Sierpinski carpet inspired fractal structure.}\label{fig2}
\end{figure}

\begin{figure}
	\centering
	\includegraphics[trim=9.7cm 0cm 0cm 0cm,width=16.6cm,clip]{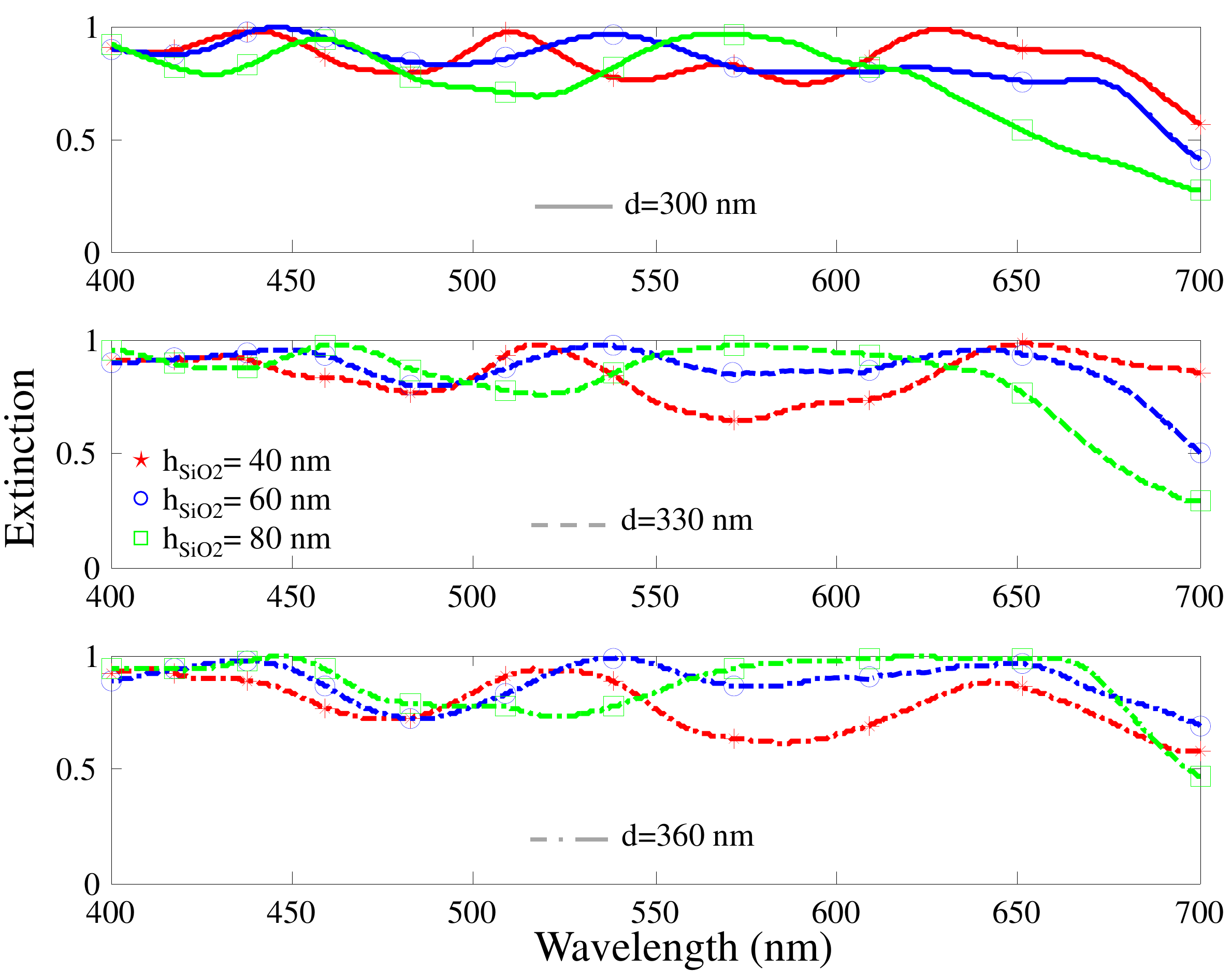}\\
	\captionsetup{justification=justified}
	\caption{Simulated extinction spectra of the proposed design as a function of lattice constant ($d$) and dielectric spacer thickness ($h_{SiO_{2}}$) at normal incidence. Thickness of the silver backreflector and the central nanobrick are chosen as $h_{Ag}=100$ nm and $h_{1}=210$ nm, respectively. }\label{fig3}
\end{figure}

Now, consider a plane wave impinges toward the structure shown in Fig.~\ref{fig1}(b) at normal incidence. Fig.~\ref{fig3} presents the optical extinction spectrum of the structure for verious values of $d$ and $h_{SiO_{2}}$ while other geometric parameters are chosen as $d2=d1/3=d/9$, $h2=h1/3=h/9=70$ nm, $h_{Ag}=100$ nm and $p=d/3$. Optical extinction is defined by $1-T(\omega)-R(\omega)$ in which $T(\omega)$ and $R(\omega)$ are frequency-dependent total transmittance and reflectance from the MDM absorber, respectively. It is notable that using an optically thick silver backreflector hinders light transmission and approximately reduces the optical extinction to $1-R(\omega)$. Obviously, the lower the amount of the $R(\omega)$ and $T(\omega)$, the higher the performance of the absorber. It should be noted that reflectance $R(\omega)$ can be readily obtained from the $S$-parameters $R(\omega)=|S_{11}(\omega)^{2}|$. Moreover, since the operating wavelength window is at least $1.3$ times larger than the lattice constant of the unit cell, no higher order diffraction is considered due to the subwavelength characterization of each unit cell.

An optimal thickness of the dielectric layer exists which maximizes the extinction. With the aforementioned geometry parameters, when $h_{SiO_{2}}=60$ nm, the reflectance has its minimum values in most wavelengths resulting in higher average absorption. For the present geometry, absorption is mainly attributed to metal losses and plasmon coupling between the silver nanocuboids and the backreflector. By increasing the thickness of the dielectric layer, coupling strength will be enhanced first until reaching its maximum value, then decreases thereafter \cite{hao2010high}. When the nanocuboids array resides far away from the silver backreflector, the optical extinction of such system is dominantly determined by the dielectric layer properties \cite{leveque2006tunable}.

\begin{figure}
	\centering
	\includegraphics[trim=10cm 0cm 0cm 0cm,width=16.6cm,clip]{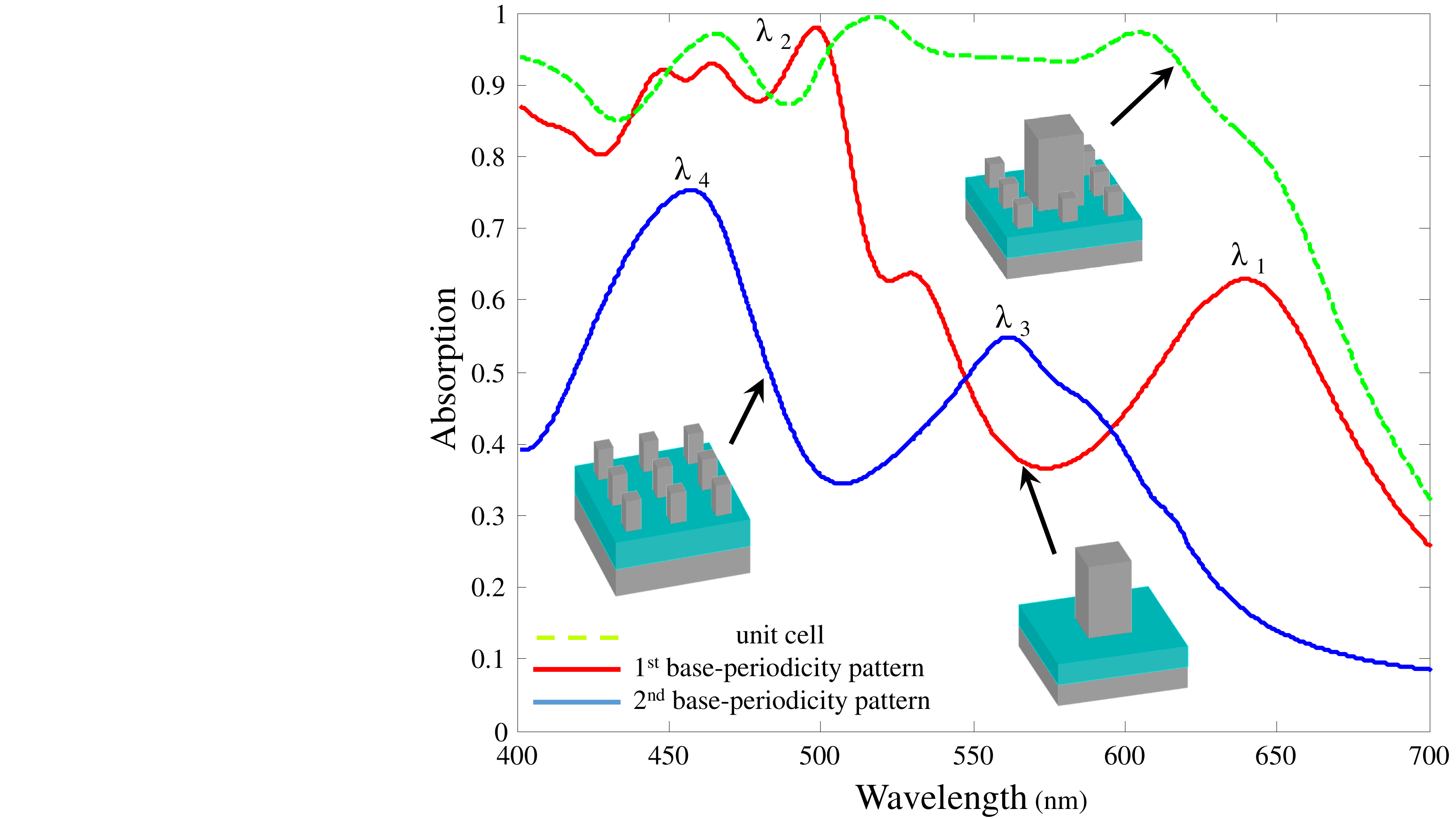}\\
	\captionsetup{justification=justified}
	\caption{ Calculated extinction spectra for the two base-periodicity patterns as well as the main unit call of the proposed absorber for $d=330$ nm, $h_{SiO_{2}}=60$, and $h=630$ nm.}\label{fig4}
\end{figure}

To investigate the electromagnetic wave trapping in the dielectric layer and plasmonic fractal, we first break down the fractal-like structure into two base-periodicity patterns as shown in Fig.~\ref{fig4}. By exploring the underlying mechanism of these two simple individual patterns, physical analysis of the main complex unit cell can readily be revealed. The corresponding calculated extinction spectra over the wavelength window ($400-700$ nm) are depicted in Fig.~\ref{fig4}. To disclose the physical origin of the absorption, the electromagnetic field distributions for these resonant modes should be first presented.
As Fig.~\ref{fig4} shows, two distinct peaks observed for the first base-periodicity pattern: one is around the wavelength $\lambda_{1}=638$ nm and the other is about  $\lambda_{2}=497$ nm. These peaks are attributed to the electric/magnetic dipolar resonances and the presence of surface plasmon modes; although, contribution of parasitic absorption of silver nanocuboids and backreflector plays an important role in enhancement of optical extinction.

\begin{figure} 
	\centering
	\includegraphics[trim=0cm 9cm 15.3cm 0cm,width=16.6cm,clip]{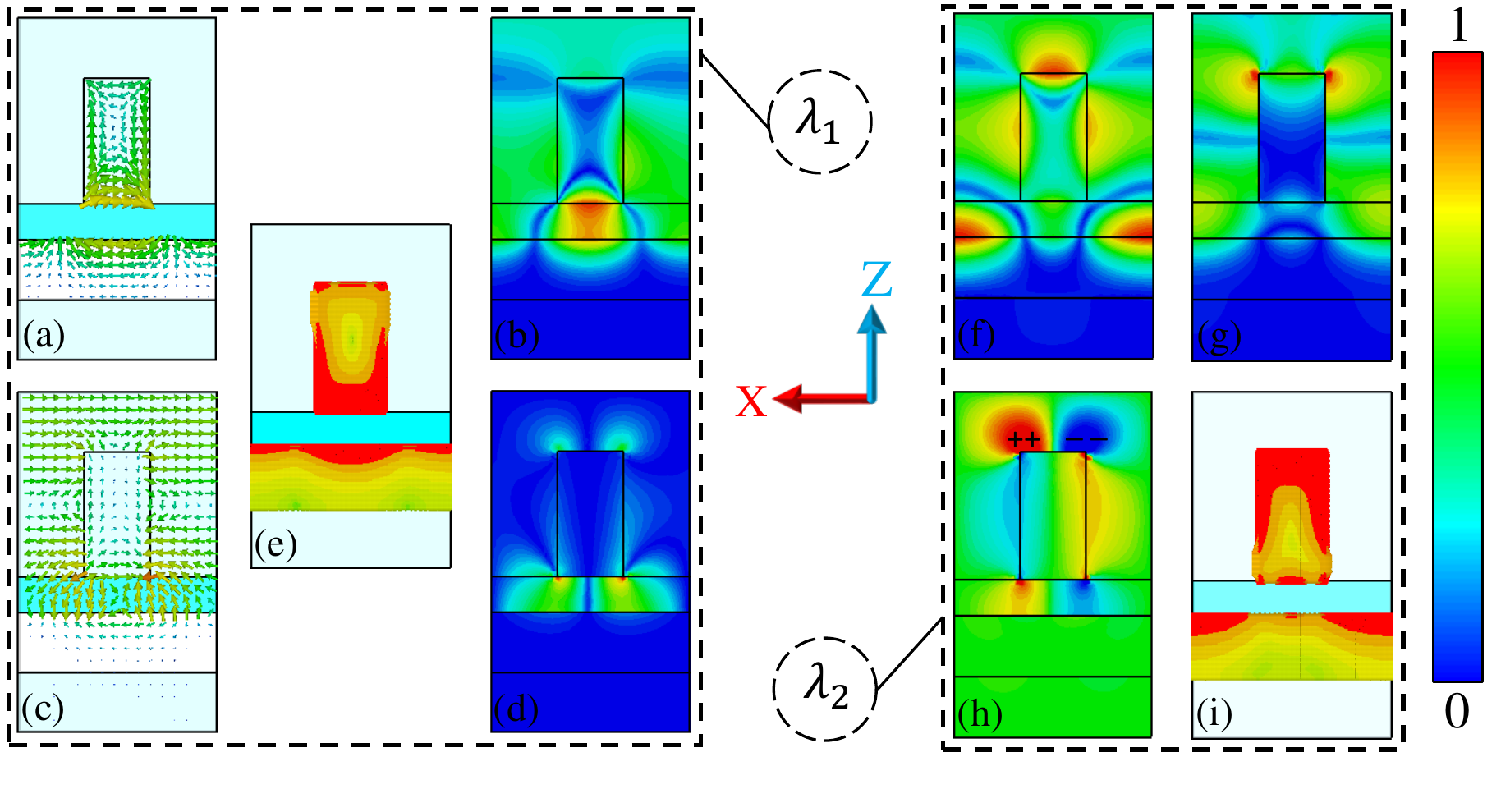}\\
	\captionsetup{justification=justified}
	\caption{Representation of (a) the electric displacement vectors, (b,f) total magnetic field intensity distribution, (c) the electric field polarization vectors, (d,g) the distributions of electric field intensity, (e,i) arrows of the time-averaged resistive loss density, and (h) real part of electric field along the $x$-$z$ plane for incidence TM-polarized light with the wavelength $\lambda_{1}=638$ nm and $\lambda_{2}=497$ nm, respectively.}\label{fig6}
\end{figure}

For the absorption peak at $\lambda_{1}$, the electric displacement vectors are represented by the arrows in both of nanocuboids and the backreflector (Fig.~\ref{fig6}(a)). This opposite electric displacement generates circulating currents between the top and bottom plasmonic layers. Such circulating current, namely magnetic resonance leads to an artificial magnetic moment that interacts significantly with the magnetic field of the incident light \cite{hao2010high}. As a result, an intense enhancement of the localized electromagnetic fields is emerged between the two plasmonic layers as shown in Fig.~\ref{fig6}(b) by surface colormap. Furthermore, the electric field distribution amplitude and  polarization plotted in Fig.~\ref{fig6}(c) and Fig.~\ref{fig6}(d) imply that the accumulated charges at the both sides of nanocuboids which couple to their image charges in the silver backreflector generate strong electric dipole resonances. As a result, such strong resonances trap and confine incident electromagnetic energy efficiently to the intermediate dielectric spacer and provide enough time to be dissipated as heat due to the Ohmic losses within the plasmonic section. The feature of enhanced absorption is further illustrated with the time-averaged resistive loss density for the absorption peaks. Fig.~\ref{fig6}(e) depicts that a large portion of incident power is absorbed by the plasmonic nanocuboids and the silver backreflector.

The absorption peak at $\lambda_{2}$ is related to the excitation of SP modes at the interface between the silicon dioxide dielectric spacer and the ground mirror due to the coupling effect of periodic top silver nanocuboids. The main advantage of SP modes is their tunability by adjusting the the shape and size of subwavelength plasmonic structures. It is worth mentioning that at larger periods, the metal nanocuboids array will excite lower-energy modes by providing smaller momenta \cite{liu2015multimodal}.
Fig.~\ref{fig6}(f) and Fig.~\ref{fig6}(g) depict that the magnetic and electric fields are mainly confined at the silicon dioxide/silver interface which confirms the typical feature of interface-bounded SP modes. Furthermore, as shown in Figs.~\ref{fig6}(h), high electric field intensity around the metal edges implies the existence of localized surface plasmon mode. Such an anti-phase electric field distribution is induced by the accumulation of opposite charges at the tips of silver nanopatterns. Due to the strong coupling of resonant electric modes excited at sharp corners of nanocuboids, electric field is significantly boosted between the gap of adjacent metallic nanoparticles. Figs.~\ref{fig6}(i) illustrates that power dissipation due to resistive loss almost occurs in the nanocubids. 

\begin{figure}[t!] 
	\centering
	\includegraphics[trim=0cm 9cm 15.3cm 0cm,width=16.6cm,clip]{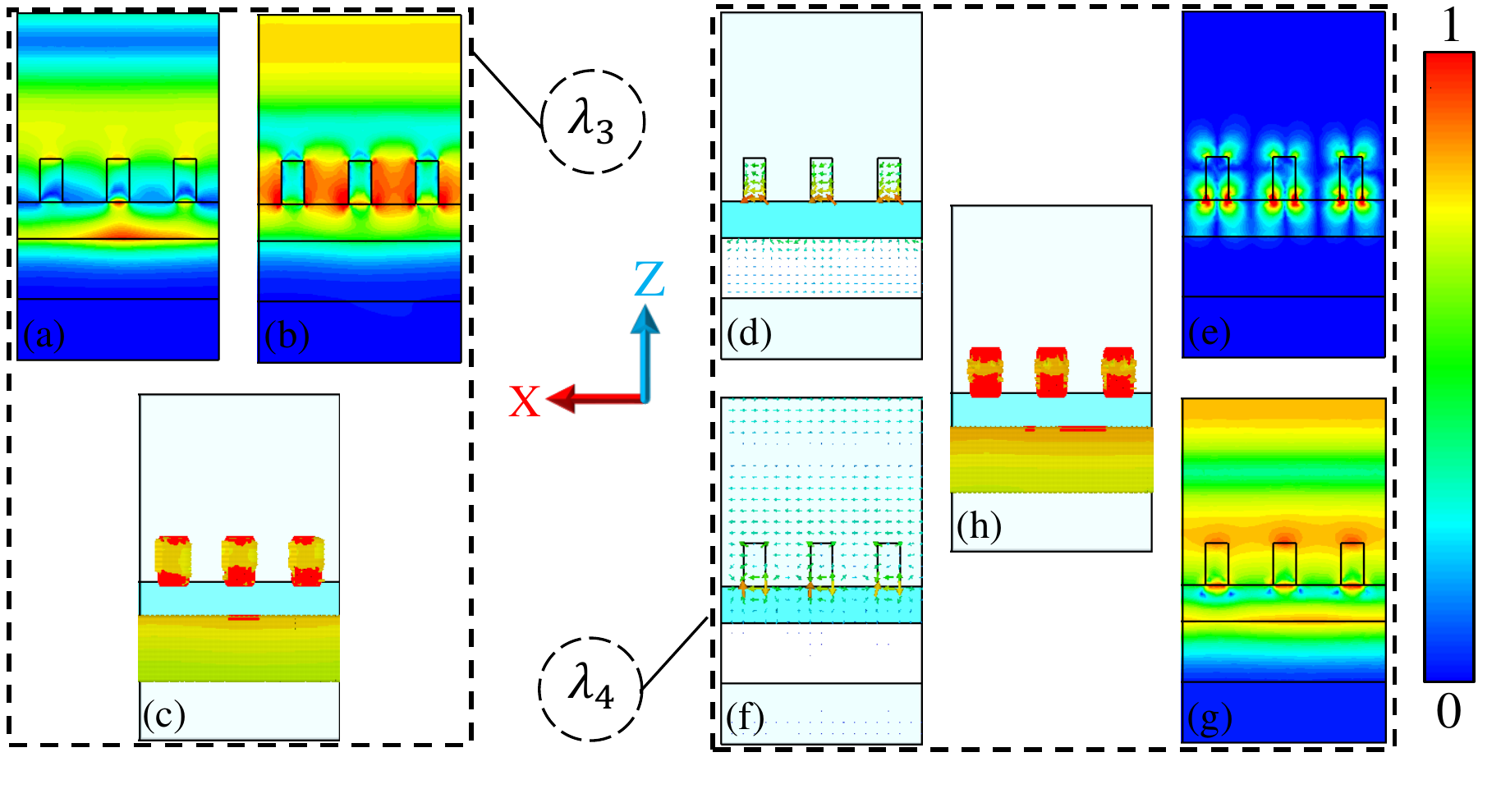}\\
	\captionsetup{justification=justified}
	\caption{Representation of (a,g) total magnetic field intensity distribution, (b,e) the distributions of electric field intensity, (c,h) arrows of the time-averaged resistive loss density, (d) the electric displacement vectors, and (f) the electric field polarization vectors along the $x$-$z$ plane for incidence TM-polarized light with the wavelength $\lambda_{3}=560$ nm and $\lambda_{4}=455$ nm, respectively.}\label{fig8}
\end{figure}

For the case of $3\times3$ small nanocuboids array shown in Fig.~\ref{fig4}, two peaks are emerged at $\lambda_{3}=560$ nm and $\lambda_{4}=455$ nm. As Fig.~\ref{fig8}(a) and Fig.~\ref{fig8}(b) depict, cavity mode which comes from the interaction between the incident light and the reflected light from silver back reflector is appeared at $\lambda_{3}$. Such a cavity mode confirms the Fabry-Perot resonance at the longitudinal cavities which is independent of the plasmonic effect and is determined by the dielectric spacer thickness \cite{zhu2013broadband}.

The physical mechanism of absorption at $\lambda_{4}$ is similar to $\lambda_{1}$; the electric field is mainly confined in the corner areas of the silver nanocuboids and the inside area of the intermediate dielectric. Fig.~\ref{fig8}(f) demonstrates the significant amount of charge in the corner areas of the top metallic section which verifies the excitation of the electric dipolar resonances. Furthermore, as Fig.~\ref{fig8}(a) depicts the anti-parallel current creating a loop between the silver nanosquares array and the bottom silver film. This loops gives rise to artificial magnetic dipole moments that interact strongly with the incident magnetic field. 

\begin{figure} 
	\centering
	\includegraphics[trim=10.8cm 9cm 0cm 0cm,width=16.6cm,clip]{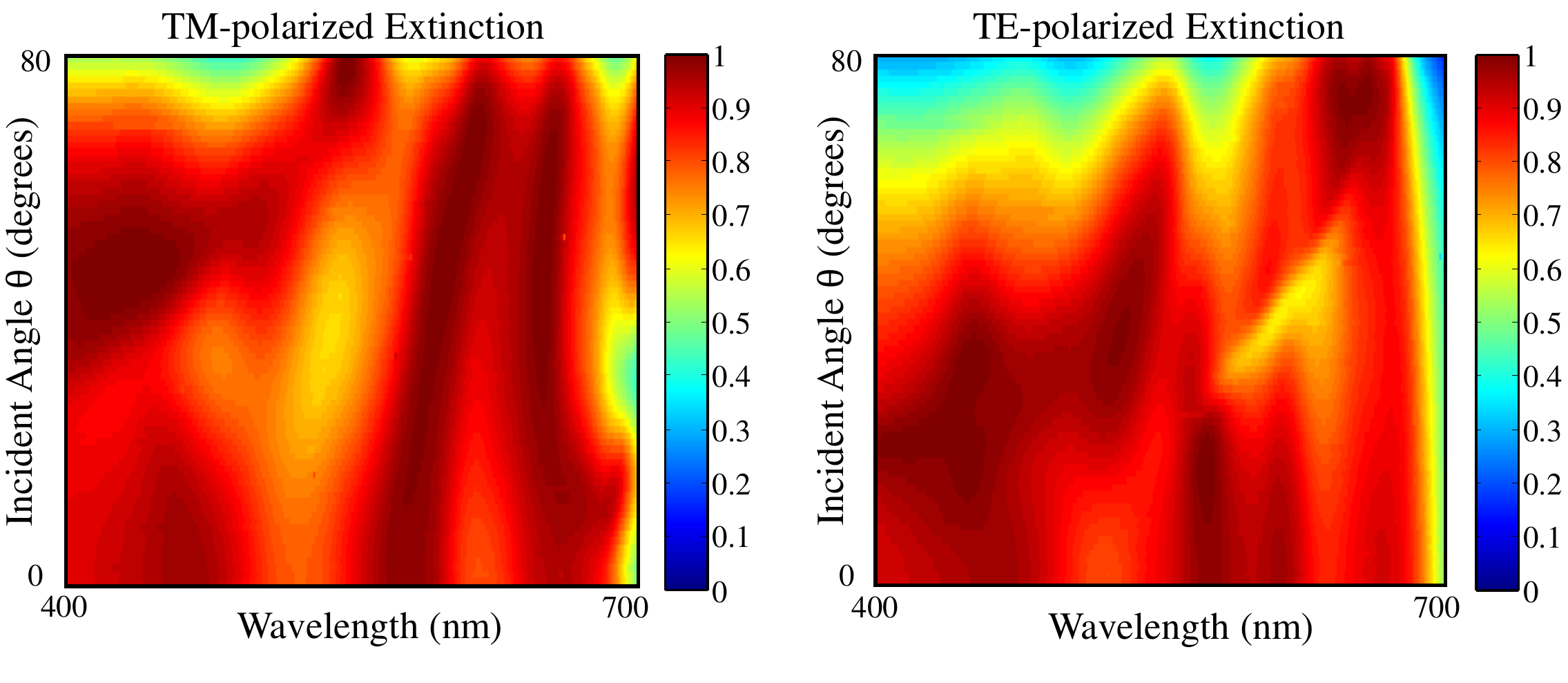}\\
	\captionsetup{justification=justified}
	\caption{Simulated extinction spectra of the proposed design versus angle of incidence of a (a) TM-polarized and (b) TE-polarized light.}\label{fig10}
\end{figure}

By revealing the fundamental mechanism of extinction in our proposed fractal-like absorber (Fig.~\ref{fig1}), now, we focus on robustness of the absorption spectrum against oblique incident angles. Since the whole structure is symmetric in $x$ and $y$ directions, we expect approximately similar responses for both transverse-magnetic (TM) and transverse-electric (TE) polarized electromagnetic waves under oblique incident. To this end, a further study of polarization and angle dependent performance of the  absorption is performed. Fig.~\ref{fig10}(a) and Fig.~\ref{fig10}(b) present the absorption behaviors of the broadband absorption under a tuning oblique incidence with TM and TE polarization states, respectively. It is observed that the proposed absorber  accomplishes absorptance higher than $80$\% in a wide angle range over almost the entire solar spectrum. Such polarization-independent and incident angle insensitive broadband light absorption is achieved due to the symmetry structural feature of fractal-like pattern and excitation of localized resonances in the system.

\section{CONCLUSION}
 
In summary, we numerically demonstrated a broadband, polarization-insensitive and wide-angle plasmonic absorber with top silver fractal-like pattern in the form of Sierpinski nanocarpets. Simulation results demonstrated average extinction of $90$\% in the spectral range of $400-700$ nm due to the self-similarity property in the plasmonic fractal. Moreover, the extinction spectra show substantial overlap between both TE and TM-polarized electromagnetic waves in a wide range of incident angles. Physical interpretation of the underlying mechanism of absorption was carried out through simulation of magnetic field, electric field, and displacement current for the two-base periodicity patterns of main structure. It was well revealed that broadband absorption profile is attributed to the excited cavity modes, SP modes, and electric/magnetic dipoles. The proposed configuration is a promising candidate for enhancing light trapping in sensing and camouflage applications.

\providecommand*\mcitethebibliography{\thebibliography}
\csname @ifundefined\endcsname{endmcitethebibliography}
  {\let\endmcitethebibliography\endthebibliography}{}


\newpage

\end{document}